\documentclass[aps,prd,twocolumn,showpacs,amsmath,amssymb]{revtex4-1}
\usepackage{amsmath}
\usepackage{graphicx}
\usepackage{subfigure}
\usepackage{epstopdf}
\usepackage{color}
\usepackage{multirow}
\usepackage{setspace}
\usepackage{overpic}
\usepackage{amssymb}
\usepackage[bookmarksnumbered, pdfstartview=FitH,colorlinks,urlcolor=blue, citecolor=blue,linkcolor=blue] {hyperref}
\usepackage{lineno}
\usepackage{bm}
\usepackage{rotating}
\usepackage[utf8]{inputenc}

\let\oldequation\equation
\let\oldendequation\endequation
\renewenvironment{equation}
  {\linenomathNonumbers\oldequation}
  {\oldendequation\endlinenomath}

\begin{document}

\title{\bf \boldmath
Determination of the absolute branching fractions of $D^0\to K^-e^+\nu_e$ and $D^+\to \bar K^0 e^+\nu_e$}

\author{
M.~Ablikim$^{1}$, M.~N.~Achasov$^{10,b}$, P.~Adlarson$^{67}$,
S. ~Ahmed$^{15}$, M.~Albrecht$^{4}$, R.~Aliberti$^{28}$,
A.~Amoroso$^{66A,66C}$, M.~R.~An$^{32}$, Q.~An$^{63,49}$,
X.~H.~Bai$^{57}$
% ...
\\
\vspace{0.2cm}
(BESIII Collaboration)\\
\vspace{0.2cm} {\it
$^{1}$ Institute of High Energy Physics, Beijing 100049, People's Republic of China\\
% ...
$^{4}$ Bochum Ruhr-University, D-44780 Bochum, Germany\\
% ...
$^{10}$ G.I. Budker Institute of Nuclear Physics SB RAS (BINP), Novosibirsk 630090, Russia\\
% ...
$^{15}$ Helmholtz Institute Mainz, Staudinger Weg 18, D-55099 Mainz, Germany\\
% ...
$^{28}$ Johannes Gutenberg University of Mainz, Johann-Joachim-Becher-Weg 45, D-55099 Mainz, Germany\\
% ...
$^{32}$ Liaoning Normal University, Dalian 116029, People's Republic of China\\
% ...
$^{49}$ State Key Laboratory of Particle Detection and Electronics, Beijing 100049, Hefei 230026, People's Republic of China\\
% ...
$^{57}$ University of Jinan, Jinan 250022, People's Republic of China\\
% ...
$^{63}$ University of Science and Technology of China, Hefei 230026, People's Republic of China\\
% ...
$^{66}$ University of Turin and INFN, (A)University of Turin, I-10125, Turin, Italy; (B)University of Eastern Piedmont, I-15121, Alessandria, Italy; (C)INFN, I-10125, Turin, Italy\\
$^{67}$ Uppsala University, Box 516, SE-75120 Uppsala, Sweden\\
% ...
$^{b}$ Also at the Novosibirsk State University, Novosibirsk, 630090, Russia\\
% ...
}
}
\vspace{0.4cm}

\begin{abstract}
Using 2.93 fb$^{-1}$ of $e^+e^-$ collision data collected with the BESIII detector at a center-of-mass energy of 3.773~GeV,
we measure the absolute branching fractions of the decays $D^0\to K^-e^+\nu_e$ and $D^+\to \bar K^0 e^+\nu_e$
to be$(3.567\pm0.031_{\rm stat}\pm 0.025_{\rm syst})\%$ and $(8.68\pm0.14_{\rm stat}\pm 0.16_{\rm syst})\%$, respectively.
Starting with the process $e^+e^-\to D\bar{D}$, a new reconstruction method is employed to select events that contain candidates for both
$D\to \bar Ke^+\nu_e$ and $\bar D\to Ke^-\bar \nu_e$ decays.
The branching fractions reported in this work are consistent within uncertainties with previous BESIII measurements that selected events containing $D\to \bar Ke^+\nu_e$ and hadronic $\bar D$ decays.
Combining our results with the lifetimes of the $D^0$ and $D^+$ mesons and the previous BESIII measurements leads to a ratio of the two decay partial widths of
$\frac{\bar \Gamma_{D^0\to K^{-}e^+\nu_{e}}}{\bar \Gamma_{D^{+}\to \bar K^{0}e^+\nu_{e}}}=1.039\pm0.021$.
This ratio supports isospin symmetry in the $D^0\to K^-e^+\nu_e$ and $D^+\to \bar K^0 e^+\nu_e$ decays within $1.9\sigma$.
\end{abstract}


\maketitle

\oddsidemargin  -0.2cm
\evensidemargin -0.2cm

\section{Introduction}
Experimental studies of semileptonic $D^{0(+)}$ decays are important for our understanding of the strong and weak interactions in charmed meson decays.
Among all exclusive semileptonic $D^{0(+)}$ decays, the $D^0\to K^-e^+\nu_e$ and $D^+\to \bar K^0e^+\nu_e$ decays have the largest branching fractions, the cleanest experimental signatures, and the highest event yields. Precise measurements of these decay branching fractions
probe nonperturbative effects in heavy meson decays, and can be used to validate the theoretical predictions~\cite{CCQM,HMchiF,cheng,LCSR} shown in Table \ref{tab:theory}.
Moreover, an accurate measurement of the ratio of branching fractions of the decays $D^0\to K^-e^+\nu_e$ and $D^+\to \bar K^0e^+\nu_e$ is an important test of isospin symmetry in the context of weak decays.
Finally, as pointed out in Ref.~\cite{epjc}, semileptonic branching fractions can be used to determine the product of the hadronic form factor $f^K_+(0)$ and the Cabibbo-Kobayashi-Maskawa (CKM) matrix element $|V_{cs}|$.  Measuring $f^K_+(0)$  tests key predictions from Lattice QCD calculations; and measuring $|V_{cs}|$ probes the unitarity of the CKM matrix.

Direct measurements of the branching fractions of the $D^0\to K^-e^+\nu_e$ and $D^+\to \bar K^0e^+\nu_e$  decays have previously been reported
by BES~\cite{bes2-kev,bes2-ksev}, Belle~\cite{k2}, CLEO~\cite{k1}, and BESIII~\cite{kev,ksoev,cpc40,k8};
and indirect measurements of the $D^0\to K^-e^+\nu_e$ decay
have previously been presented by E691~\cite{k6}, CLEO~\cite{k4,k5}, and BaBar~\cite{k3}.
In this work, a new technique is used to determine the absolute branching fractions of the decays $D^0\to K^-e^+\nu_e$  and $D^+\to \bar K^0e^+\nu_e$ by reconstructing both
$D\to \bar Ke^+\nu_e$ and $\bar D\to Ke^-\bar \nu_e$ within the same events.
We use 2.93 fb$^{-1}$ of $e^+e^-$ collision data collected with the BESIII detector at a center-of-mass energy of $\sqrt s=3.773$~GeV.

\begin{table*}[htp]
\centering
\caption{\small
Branching fractions (in \%) of $D^0\to K^-e^+\nu_e$ and $D^+\to \bar K^0e^+\nu_e$ predicted by the covariant confined quark model (CCQM)~\cite{CCQM}, chiral unitary approach (HM$\chi$T)~\cite{HMchiF}, light-front quark model (LFQM)~\cite{cheng}, and light-cone QCD sum rules (LCSR)~\cite{LCSR}, as well as comparison with the CLEO, BESIII, and world average values.}
\small
\begin{tabular}{cccccccc}
  \hline
  \hline
Decay &CCQM~\cite{CCQM}&HM$\chi$T~\cite{HMchiF}&LFQM~\cite{cheng}&LCSR~\cite{LCSR}&CLEO-c~\cite{k1}&BESIII~\cite{kev}~\cite{ksoev}&PDG~\cite{pdg2020} \\ \hline
$D^0\to K^-e^+\nu_e$     &3.63&3.4&--&$3.20^{+0.47}_{-0.43}$&$3.50\pm0.03\pm0.04$&$3.505\pm0.014\pm0.033$&$3.542\pm0.035$\\
$D^+\to \bar K^0e^+\nu_e$&9.28&8.4&$10.32\pm0.93$&$8.12^{+1.19}_{-1.08}$&$8.83\pm0.10\pm0.20$&$8.60\pm0.06\pm0.15$&$8.73\pm0.10$\\
  \hline
  \hline
\end{tabular}
\label{tab:theory}
\end{table*}

\section{Measurement method}

The $\psi(3770)$ resonance decays predominately into $D\bar{D}$ meson pairs and
our new method finds candidate events that either include
$D^0\to K^-e^+\nu_e$ and $\bar D^0\to K^+e^-\bar \nu_e$ decays or
$D^+\to \bar K^0e^+\nu_e$ and $D^-\to K^0e^-\bar \nu_e$ decays.  These are called neutral and charged double-tag~(DT) events, respectively.
The yield of the DT signal events is given by
\begin{equation}
N_{\rm DT} =  N_{D\bar{D}}\cdot{\mathcal B}^{2}_{\rm SL}\cdot \epsilon_{\rm DT},
\end{equation}
where $N_{D\bar{D}}$ is the total number of $D^0\bar D^0$ or $D^+D^{-}$ pairs in the data set,
${\mathcal B}_{\rm SL}$ is the branching fraction of the signal $D^0\to K^-e^+\nu_e$ or $D^+\to \bar K^0e^+\nu_e$ decay; and
$\epsilon_{\rm DT}$ is the efficiency of detecting the DT signal events.
The branching fraction of the $D^0\to K^-e^+\nu_e$ or $D^+\to \bar K^0e^+\nu_e$ decay can then be determined by
\begin{equation}
\label{eq:br}
{\mathcal B}_{\rm SL} = \sqrt{N_{\rm DT}/(N_{D\bar{D}}\cdot \epsilon_{\rm DT})}.
\end{equation}
In our previous work, we have determined $N_{D^0\bar D^0}=(10597\pm28\pm98)\times 10^3$ and $N_{D^+D^-}=(8296\pm31\pm65)\times 10^3$ using hadronic $D$ decays~\cite{NDD}.
Therefore, the semileptonic signal events used in this analysis are independent of the quoted numbers of $N_{D^0\bar D^0}$ and $N_{D^+D^-}$.

\section{BESIII detector and Monte Carlo}

The BESIII detector is a magnetic
spectrometer~\cite{BESIII} located at the Beijing Electron
Positron Collider (BEPCII)~\cite{Yu:IPAC2016-TUYA01,cpc41}. The
cylindrical core of the BESIII detector consists of a helium-based
 multilayer drift chamber (MDC), a plastic scintillator time-of-flight
system (TOF), and a CsI(Tl) electromagnetic calorimeter~(EMC),
which are all enclosed in a superconducting solenoidal magnet
providing a 1.0~T magnetic field. The solenoid is supported by an
octagonal flux-return yoke with resistive plate counter muon
identifier modules interleaved with steel. The acceptance of
charged particles and photons is 93\% over the $4\pi$ solid angle. The
charged-particle momentum resolution at $1~{\rm GeV}/c$ is
$0.5\%$, and the specific energy loss ($dE/dx$) resolution is $6\%$ for the electrons
from Bhabha scattering. The EMC measures photon energies with a
resolution of $2.5\%$ ($5\%$) at $1$~GeV in the barrel (end cap)
region. The time resolution of the TOF barrel part is 68~ps, while
that of the end cap part is 110~ps.

Simulated event samples, produced with the {\sc geant4}-based~\cite{geant4} Monte Carlo (MC) package and including the geometric description of the BESIII detector and the
detector response, are used to determine the detection efficiency
and estimate the backgrounds. The simulation includes the beam-energy spread and initial-state radiation (ISR) in the $e^+e^-$
annihilations modeled with the generator {\sc kkmc}~\cite{kkmc}.
The inclusive MC samples consist of the production of $D\bar{D}$
pairs with consideration of quantum coherence for all neutral $D$
modes, the non-$D\bar{D}$ decays of the $\psi(3770)$, the ISR
production of the $J/\psi$ and $\psi(3686)$ states, and the
continuum processes.
The known decay modes are modeled with {\sc
evtgen}~\cite{evtgen} using the branching fractions taken from the
Particle Data Group (PDG)~\cite{pdg2020}, and the remaining unknown decays
from the charmonium states are modeled with {\sc
lundcharm}~\cite{lundcharm}.
Final state radiation
from charged final state particles is incorporated with the {\sc
photos} package~\cite{photos}.
The $D^0\to K^-e^+\nu_e$ and $D^+\to \bar K^0 e^+\nu_e$ decays are simulated with the modified pole model~\cite{MPM}
with the pole mass fixed to the $D_s^{*+}$ nominal mass~\cite{pdg2020} and the other parameters taken from our measurements in Refs.~\cite{kev} and \cite{ksoev}, respectively.

\section{Event selection}

The selection criteria of charged and neutral kaons as well as electrons ($e^+$ or $e^-$) are the same as those used in Refs.~\cite{epjc76,cpc40,bes3-pimuv,bes3-Dp-K1ev,bes3-etaetapi,bes3-omegamuv,bes3-etamuv,bes3-etaX,bes3-DCS-Dp-K3pi,bes3-D-KKpipi,bes3-D-b1enu}.
All charged tracks are required to be within a polar angle ($\theta$) region of $|\cos\theta|<0.93$.
All charged tracks not from $K^0_S$ decays
are required to originate from the interaction point with a distance of closest approach less than 1~cm in the transverse plane perpendicular to the MDC axis and less than 10~cm along the MDC axis.

Particle identification (PID) of charged kaons is performed by combining $dE/dx$ and TOF information. For electron candidates, EMC information is also incorporated.
Charged tracks satisfying confidence levels $CL_e>$ 0.001 and $CL_e/(CL_e+CL_\pi+CL_K)>0.8$ are assigned as electron candidates.
Kaon candidates are required to satisfy $CL_{K}>CL_{\pi}$,
but must fail to satisfy the electron PID requirements.
This reduces background from Bhabha scattering events with a radiative photon converting into an $e^+e^-$ pair. To suppress misidentification between electrons and hadrons, the ratio of the deposited energy in the EMC of
an electron candidate to its momentum in the MDC is required to be within~(0.8,\,1.2).

Neutral kaon candidates are reconstructed by using the $\bar K^0\to K^0_S\to\pi^+\pi^-$ decay mode.
For two charged pion candidates, the distance of the closest approach to the interaction point is required to be less than 20~cm along the MDC axis. They are assigned as $\pi^+\pi^-$ without PID requirements. The two charged pion candidates are constrained to originate from a common vertex and are required to have an invariant mass within $(0.486,0.510)~{\rm GeV}/c^2$,
which corresponds to about three times the resolution.
The decay length of the $K^0_S$ candidate is required to have a decay vertex more than two standard deviations away from the interaction point.

Electrons often produce final state radiation and bremsstrahlung (collectively named FSR in the following) in the innermost parts of the detector.
We identify FSR photons as EMC showers that occur within $15^\circ$ of the electron's initial direction (before curving in the magnetic field).
We further require the FSR photons deposit greater than 0.025~GeV in the barrel region of the EMC or greater than 0.050~GeV in the end cap region, and that their shower time is within 700 ns of the event start time.
The four momenta of these FSR candidates are then added back to the electron tracks.
This is called FSR recovery.
To suppress backgrounds associated with a photon converting into an $e^+e^-$ pair,
especially from the radiative Bhabha events for neutral DT candidates,
the opening angles between electron and positron candidates and between $K^+$ and $K^-$ candidates are required to satisfy $\cos\theta_{e^+e^-}<0.95$ and $\cos\theta_{K^+K^-}<0.95$.

\begin{figure*}[htbp]
\begin{center}
\includegraphics[width=0.49\linewidth]{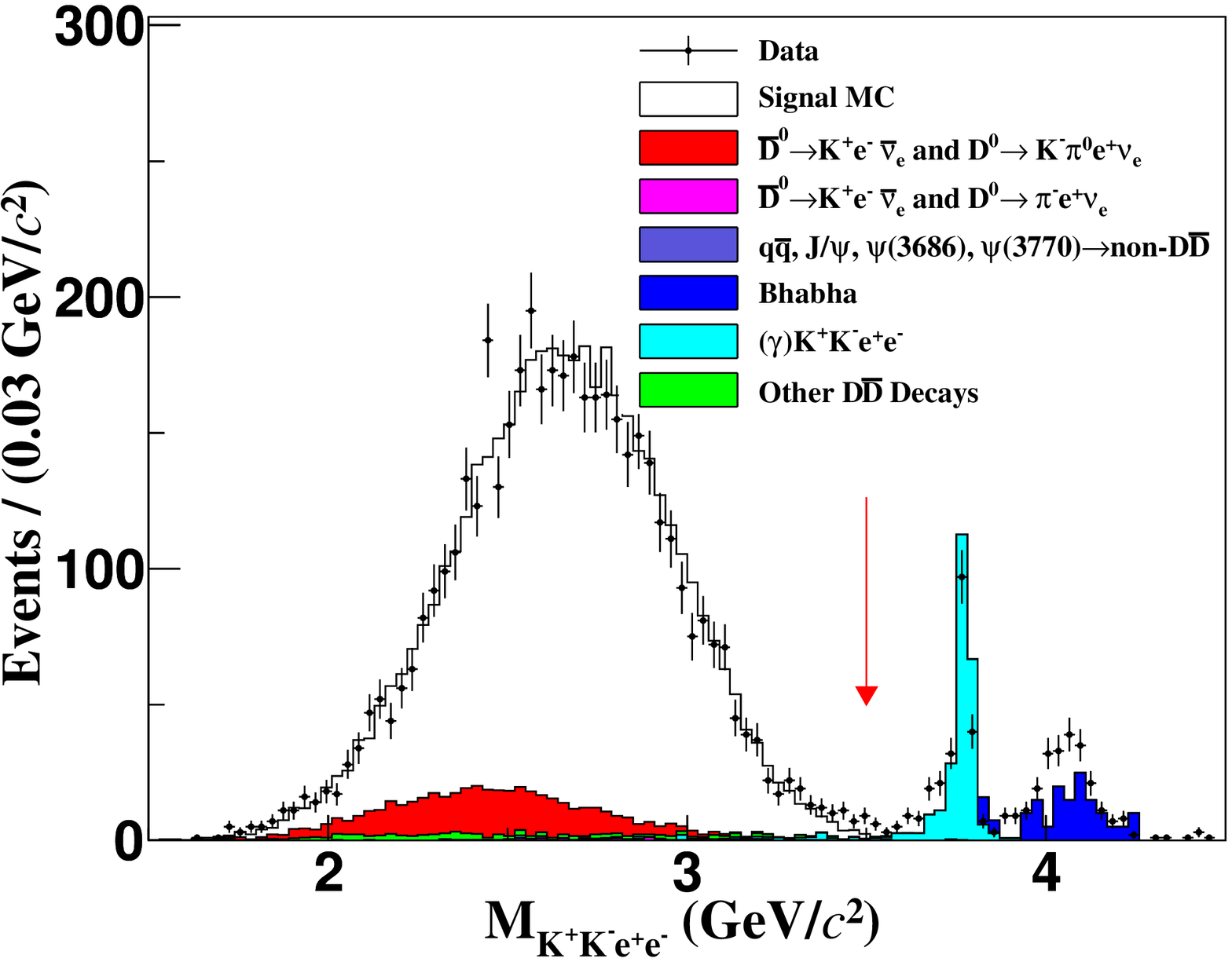}
\includegraphics[width=0.49\linewidth]{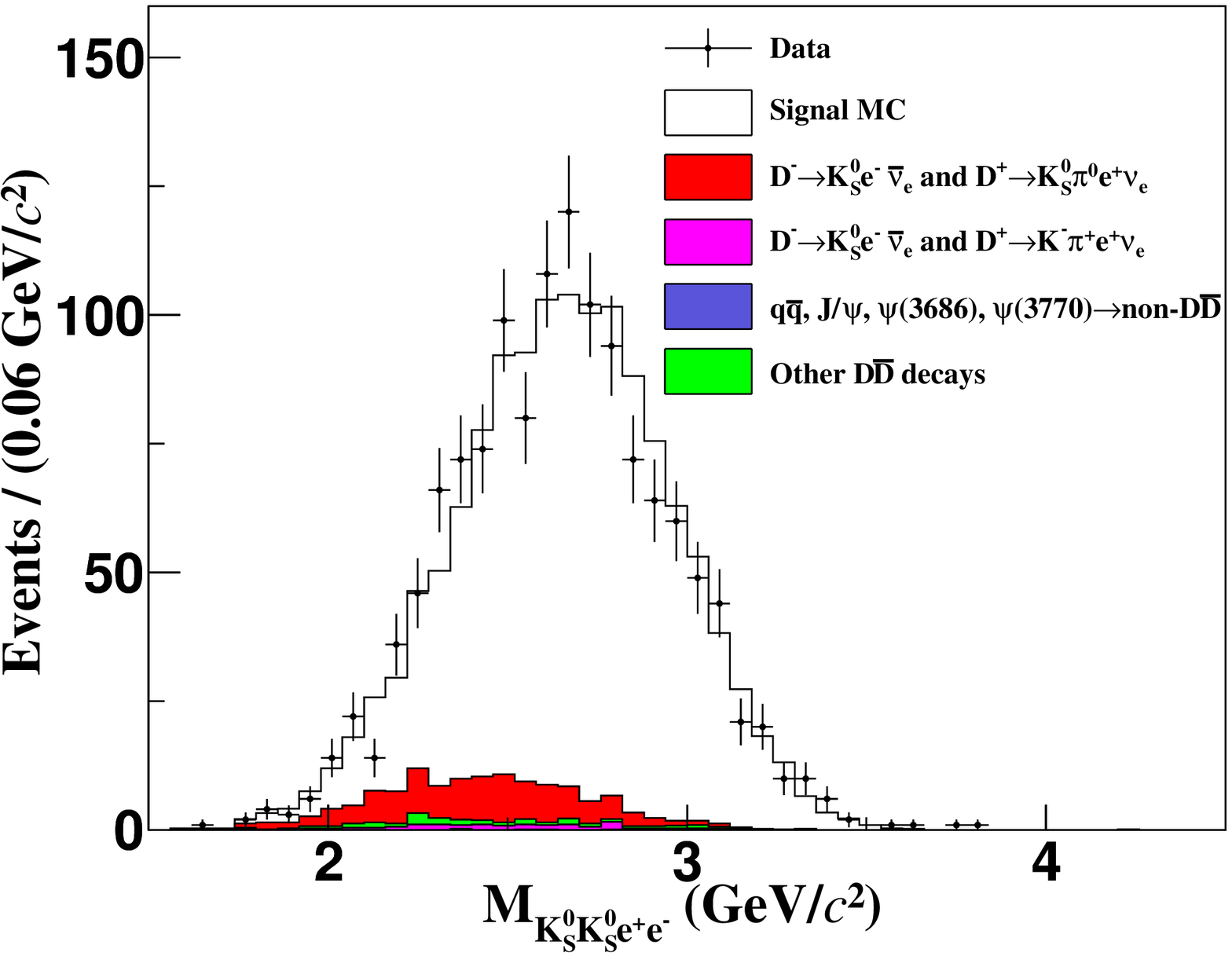}
\end{center}
\caption{
The $M_{K\bar K e^+e^-}$ distributions of the candidate events for (left) $D^0\to K^-e^+\nu_e$ and $\bar D^0\to K^+e^-\bar \nu_e$ before imposing the $M_{K^{+}K^{-}e^{+}e^{-}}$ requirement (shown as the red arrow at $3.5$~GeV/$c^2$) and
(right) $D^+\to \bar K^0e^+\nu_e$ and $D^-\to K^0 e^-\bar \nu_e$ in data (points with error bars) and the inclusive MC sample (histograms).
}
\label{fig:bbb}
\end{figure*}

Figure~\ref{fig:bbb} shows the $M_{K\bar K e^+e^-}$ distributions of the candidate events selected from data and the inclusive MC sample.
To further reject backgrounds from Bhabha scattering events
and $e^+e^-\to (\gamma)K^+K^-e^+e^-$ events,
which concentrate around 4.05 and 3.773~GeV/$c^2$ in the $M_{K^{+}K^{-}e^{+}e^{-}}$ distribution, respectively,
$M_{K^{+}K^{-}e^{+}e^{-}}$ is required to be less than 3.50~GeV/$c^2$.

The main source of background for the neutral DT channel are events with $D^0\to K^-\pi^0e^+\nu_e$ and $\bar D^0\to K^+e^-\bar \nu_e$.  For the charged DT channel, the main background comes from events with $D^+\to K^0_S\pi^0 e^+\nu_e$ and $D^-\to K_S^0e^-\bar \nu_e$.  The processes
$D^0\to K^-\pi^0e^+\nu_e$ and  $D^+\to K^0_S\pi^0 e^+\nu_e$
are dominated by $D^0\to K^{*}(892)^-e^+\nu_e$ and $D^+\to \bar K^{*}(892)^0e^+\nu_e$, respectively.
To suppress background processes with extra photon(s) and $\pi^0$(s), we require that the maximum energy of any extra photon~($E^{\rm max}_{\rm extra\,\gamma}$) is less than 0.25~GeV and there is no extra $\pi^0$ candidate in the event ($N_{\rm extra\,\pi^{0}}$).
The $\pi^0$ candidates are reconstructed via the $\pi^0\to\gamma\gamma$ decay and
the opening angle between the photon candidate and the nearest charged track is required to be greater than $10^{\circ}$.
Any photon pair with an invariant mass between $(0.115,\,0.150)$\,GeV$/c^{2}$ is regarded as a $\pi^0$ candidate, and a kinematic fit is imposed on the photon pair to constrain its invariant mass to the $\pi^0$ nominal mass~\cite{pdg2020}.

\section{Data analysis}

The numbers of DT candidate events for the neutral and charged channels are measured using the missing mass squared:
\begin{equation}
M_{\rm miss}^{2}\equiv {E_{\rm miss}^2}-{|\vec{p}_{\rm miss}|^2},
\nonumber
\end{equation}
where $E_{\rm miss}$
and $\vec{p}_{\rm miss}$ are the missing energy and momentum in the $e^+e^-$ center-of-mass frame, respectively.  They are calculated using:
\begin{equation}
E_{\rm miss}=2E_{\rm beam}-\sum\limits_i E_i,
\nonumber
 \end{equation}
 and
\begin{equation}
|\vec{p}_{\rm miss}|=\left|\sum\limits_i {\vec p}_i\right|,
\nonumber
 \end{equation}
where $E_{\rm beam}$ is the beam energy and $E_i$ and ${\vec p}_i$
are the measured energy and momentum of particle $i$ in the $e^+e^-$ center-of-mass frame, summing over $K$, $\bar K$, $e^+$, and $e^-$.
The $M_{\rm miss}^2$ distributions of the candidate events surviving in data and the inclusive MC sample are shown in Fig.~\ref{fig:opt6}.
Good consistency between data and the MC simulation can be seen.
Signal events appear in a wide range, $(-0.02,2.50)$~GeV$^2$/$c^4$, due to the two missing neutrinos in the final state.
The number of events in the signal region for neutral and charged DT candidate events
are $4449$ and $1317$, respectively.

\section{Background analysis}

Background events are divided into three categories. The first category comes from $\psi(3770)\to D\bar D$ decays.
 The corresponding background yields are estimated from the number of $D^0\bar D^0$ or $D^+D^-$ pairs in the data, the relevant branching fractions for the $D$ and $\bar D$ decays, and the individual MC-determined misidentification rates.
The branching fractions used for $D^+\to K^0_S \ell^+\nu_\ell$~($\ell=e$ or $\mu$) and $D^+\to K^0_L \ell^+\nu_\ell$ are taken as half the world average value branching fraction for $D^+\to \bar K^0 \ell^+\nu_\ell$. The $D^0\to K^-\pi^0e^+\nu_e$ and $D^0\to K_L^0\pi^- e^+ \nu_e$ decay branching fractions have not been (well) measured
so we use half of the world average value branching fraction for $D^0\to \bar K^0\pi^- e^+ \nu_e$ (dominated by $D^0\to K^*(892)^-e^+\nu_e$) using isospin symmetry.
The $D^+\to K^0_S\pi^0 e^+\nu_e$ and $D^+\to K^0_L\pi^0 e^+\nu_e$ decays also have poor or no information in the PDG~\cite{pdg2020}, so we use one quarter of the world average value
of the branching fraction for $D^+\to K^-\pi^+e^+ \nu_e$ (dominated by $D^+\to \bar K^*(892)^0e^+\nu_e$), using isospin symmetry.
For the other $D$ decays, the branching fractions used are directly taken from the PDG~\cite{pdg2020}.
For neutral DT candidates, the two largest  $D\bar D$ background sources are from events with
$D^0\to K^-\pi^0e^+\nu_e$ and $\bar D^0\to K^+e^-\bar \nu_e$+c.c. (79.1\%) and events with
$D^0\to K^-\pi^0e^+\nu_e$ and $\bar D^0\to \pi^+e^-\bar \nu_e$+c.c. (3.7\%).
For charged DT candidates, the largest two  $D\bar D$  background sources are from events with
$D^+\to K_S^0\pi^0e^+\nu_e$ and $D^-\to K_S^0e^-\bar \nu_e$+c.c. (79.7\%) and events with $D^+\to K_S^0e^+\nu_e$ and $D^-\to K^+\pi^-e^-\bar \nu_e$+c.c. (9.7\%). For each of the other sources, the background yield constitutes no more than 2.7\% of the total $D\bar D$ background yield.

The second category of backgrounds comes from
the processes $e^+e^-\to q\bar q$ ($q=u,d,s$),
ISR production of $J/\psi$ and $\psi(3686)$,
$\psi(3770)\to{\text{non-}}D\bar D$ decays, and
the processes $e^+e^-\to (\gamma)\ell^+\ell^-$ ($\ell=e$, $\mu$, $\tau$).
The numbers of these background events are determined by using
the integrated luminosity,
the observed cross section of the background processes at $\sqrt s=3.773$~GeV, and
the misidentification rates based on the MC simulation.

The third category, which is only present in neutral DT events, is from the $e^+e^-\to (\gamma)K^+K^-e^+e^-$ process.
Since the cross section for $e^+e^-\to (\gamma)K^+K^-e^+e^-$  is not well known,
 we estimate the background yield in the mass region $M_{K^+K^-e^+e^-}\in(0,3.50)\,{\rm GeV}/c^2$  by using the number of events in the data with $M_{K^+K^-e^+e^-}\in(3.50,3.90)\,{\rm GeV}/c^2$ and the MC-determined ratio of the background yields between these two regions.

\begin{table*}[htbp]
\centering
\caption{
Background sources and the numbers of estimated background events for the neutral and charged DT channels.}
The listed $e^+e^-\to \psi(3770)\to D\bar D$ decays include the charge conjugate modes.
\begin{tabular}{ccc}
  \hline
  \hline
Source & $D^0\to K^-e^+\nu_e$ and $\bar D^0\to K^+e^-\bar \nu_e$ & $D^+\to \bar K^0e^+\nu_e$ and $D^-\to K^0e^-\bar \nu_e$ \\ \hline

$D^0\to K^-e^+\nu_e$ and $\bar D^0\to K^+\pi^0e^-\bar \nu_e$                                      &$309.7\pm21.9$&- \\
$D^0\to K^-e^+\nu_e$ and $\bar D^0\to \pi^+e^-\bar \nu_e$                                         &$14.6\pm3.6$&-\\
$D^+\to \bar K^0e^+\nu_e$ and $D^-\to K^0\pi^0e^-\bar \nu_e$                                      &-&$115.1\pm10.6$\\
$D^+\to \bar K^0e^+\nu_e$ and $D^-\to K^+\pi^-e^-\bar \nu_e$                                      &-&$14.0\pm3.6$\\
$\psi(3770)\to$ other $D\bar D$ decays                       &$67.0\pm2.9$&$15.4\pm1.2$ \\
$(\gamma)K^+K^-e^+e^-$                                       &$14.2\pm2.7$  & -         \\
$q\bar q$                                                    &$1.7\pm0.5$   & $0.1\pm0.1$ \\
$(\gamma_{\rm ISR})J/\psi$ and $(\gamma_{\rm ISR})\psi(3686)$&$1.3\pm0.4$   & $0.3\pm0.2$ \\
$\psi(3770)\to{\text{non-}}D\bar D$ decays                   &$0.4 \pm0.2$  & $0.1\pm0.1$ \\
$(\gamma)\ell^+\ell^-$                                       &Negligible    & Negligible  \\
  \hline
  Total &$408.9\pm22.6$ & $145.0\pm11.3$\\
  \hline
  \hline
\end{tabular}
\label{result1}
\end{table*}

The sources  and the estimated yields of the various background modes for the neutral and charged DT events are summarized in Table~\ref{result1}.
The uncertainties in the background yields include statistical and systematic uncertainties added in quadrature. The systematic uncertainties considered include the uncertainties on the number of $D^0\bar D^0$ or $D^+D^-$ pairs and the external branching fractions (or the cross section and the integrated luminosity), as well as the tracking and PID efficiencies.
The total background yields are $409\pm23$ and $145\pm11$ for the neutral and charged DT events, respectively.

\begin{figure*}[htbp]
\begin{center}
\includegraphics[width=0.49\linewidth]{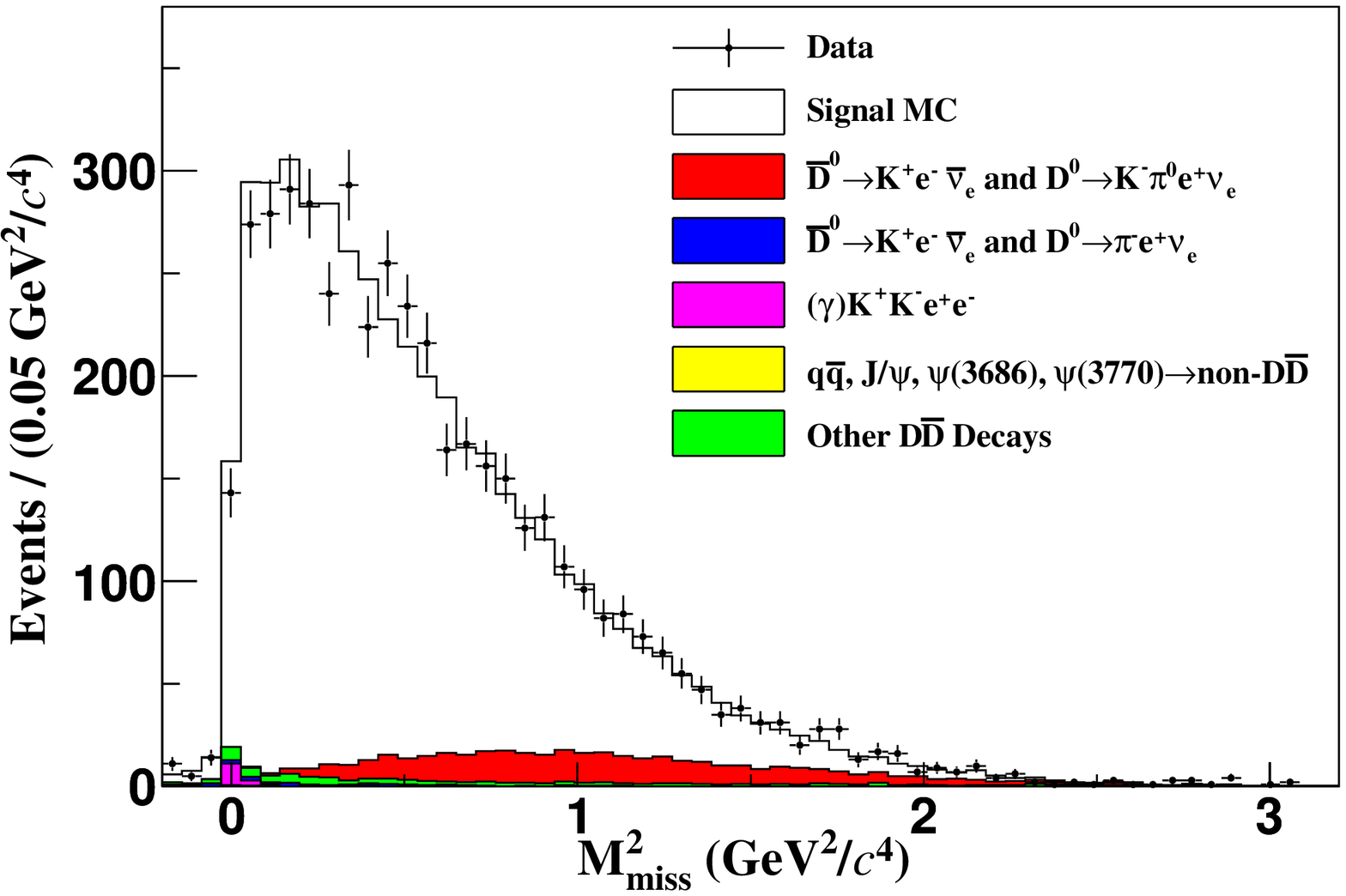}
\includegraphics[width=0.49\linewidth]{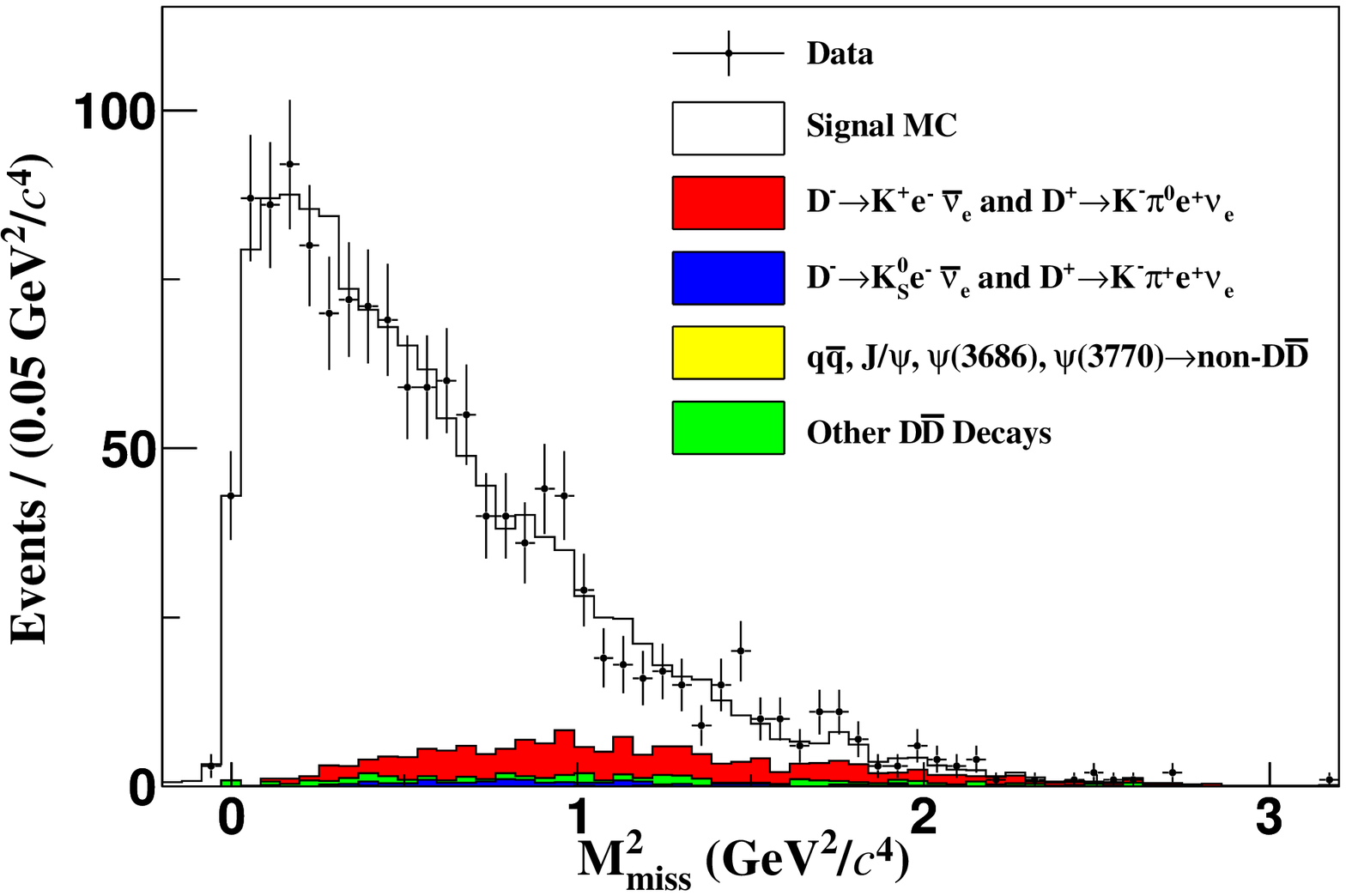}
\label{fig:opt6}
\caption{\small
The $M_{\rm miss}^2$ distributions of the (left) $D^0\to K^-e^+\nu_e$ and $\bar D^0\to K^+e^-\bar \nu_e$ and
(right) $D^+\to \bar K^0e^+\nu_e$ and $D^-\to K^0 e^-\bar \nu_e$ candidate events in data (points with error bars) and the inclusive MC sample (histograms).
\label{fig:opt6}
}
\end{center}
\end{figure*}

\section{Branching fractions}

Subtracting the background yields from the numbers of candidate events in the data,
the yields of the neutral and charged DT events ($N_{\rm DT}$) become $4040\pm70$ and $1172\pm38$, respectively.

The efficiencies of finding the neutral and charged DT signal events ($\epsilon_{\rm DT}$) are  $(29.97\pm0.03)\%$ and $(15.67\pm0.03)\%$, respectively.
For the branching fraction determination, the efficiency for $D^+$ decay needs to be corrected by the $\bar K^0\to K^0_S\to\pi^+\pi^-$~\cite{pdg2020} decay branching fraction squared.
To guarantee the reliability of the signal efficiencies,
the momentum and $\cos\theta$ ($\theta$ is the polar angle in the $e^+e^-$ rest frame) distributions of $K$, $\bar K$, $e^+$, and $e^-$ of the accepted candidate events were examined.
The data distributions are found to be well modeled by the MC simulation~\cite{kev,ksoev}.

Inserting the numbers for $N_{\rm DT}$, $N_{D\bar{D}}$, and $\epsilon_{\rm DT}$ into Eq.~(\ref{eq:br}) yields
\begin{equation}
{\mathcal B}_{D^0\to K^- e^+\nu_e}=(3.567\pm0.031_{\rm stat})\% \nonumber
\end{equation}
and
\begin{equation}
{\mathcal B}_{D^+\to \bar K^0e^+\nu_e}=(8.68\pm0.14_{\rm stat})\%. \nonumber
\end{equation}

\section{Systematic uncertainties}

The sources of the systematic uncertainties in the $D^{0}\to K^{-}e^{+} \nu_{e}$ and $D^{+}\to \bar K^{0}e^{+} \nu_{e}$ branching fraction measurements
are summarized in Table~\ref{sys1}. They are assigned relative to the measured branching fractions and are discussed below.

The uncertainties in the numbers of $D^0\bar D^0$ and $D^+D^-$ pairs in the data are 0.9\% and 0.8\%~\cite{NDD}, respectively.
Using error propagation, their effects on the measured branching fractions of $D^{0}\to K^{-}e^{+} \nu_{e}$ and $D^{+}\to \bar K^{0}e^{+} \nu_{e}$ are 0.45\% and 0.40\%, respectively.

The tracking (PID) efficiencies of $e^\pm$ and $K^\pm$ are studied with $e^+e^-\to\gamma e^+e^-$ events and DT hadronic $D\bar D$ events, respectively.
A small difference between the tracking (PID) efficiency of
data and MC simulation (called the data/MC difference) is observed.
The averaged data/MC differences of $e^\pm$ tracking and PID efficiencies, weighted by the two-dimensional (momentum and $\cos\theta$) distributions of signal MC events, are $(0.0\pm0.2)\%$ and $(-0.7\pm0.2)\%$, respectively.
The averaged data/MC differences of $K^\pm$ tracking and PID efficiencies, weighted by the corresponding momentum spectra from signal MC events, are $(0.8\pm0.2)\%$ and $(0.5\pm0.2)\%$, respectively.
After correcting the MC efficiencies for these averaged data/MC differences, the systematic uncertainties of tracking and PID efficiencies per $e^\pm$ ($K^\pm$) are assigned as 0.2\%~(0.2\%) and 0.2\%~(0.2\%), respectively.

The efficiency of $K_{S}^{0}$ reconstruction was estimated by the control samples of $J/\psi\to K^*(892)^{\mp}K^{\pm}$ ($K^*(892)^{\pm} \to K_S^{0}\pi^{\pm}$) and $J/\psi\to \phi K_S^{0}K^{\pm}\pi^{\mp}$~\cite{k8}. The difference of $K^0_S$ reconstruction efficiencies between data and MC simulation, which has been weighted by the $K_S^0$ momentum of the $D^+\to \bar K^0 e^+\nu_e$ decay, is determined to be 1.6\% per $K^0_S$ and is taken as the associated systematic uncertainty.
The uncertainties of the $E^{\rm max}_{\rm extra\,\gamma}$ and $N_{\rm extra\,\pi^{0}}$ requirements are estimated to be 0.3\% and 0.6\% for the neutral and charged DT events, respectively. These are obtained  by analyzing DT hadronic $D\bar D$ events of $D^0\to K^-\pi^+$ and $\bar D^0\to K^+e^-\bar \nu_e$ and  $D^+\to K^-\pi^+\pi^+$ and $D^-\to K^0_Se^-\bar \nu_e$.

The uncertainty of the quoted branching fraction of $K^0_S\to \pi^+\pi^-$ is 0.1\%.
The uncertainty in the MC model due to $D\to \bar Ke^+\nu_e$ is estimated to be 0.1\% by varying the form factor parameters by $\pm 1\sigma$~\cite{kev,ksoev}.
The impact of the MC model of $D\to \bar K\pi^0 e^+\nu_e$ on the branching fraction measurement is assigned as 0.1\% by varying the form factor parameters by $\pm 1\sigma$~\cite{D0_kpiev,Dp_kpiev}.
The limited numbers of MC events give uncertainties of 0.1\% and 0.2\% for $D^0\to K^-e^+\nu_e$ and $D^+\to \bar K^0e^+\nu_e$, respectively.

The uncertainties due to the requirements of ${\cos\theta}_{K^+K^-}$, ${\cos\theta}_{e^+e^-}$, $M_{K^+K^-e^+e^-}$, and $M^{2}_{\rm miss}$
are examined by choosing the alternative requirements of ${\cos\theta}_{K^+K^-}<0.93$ or 0.98, ${\cos\theta}_{e^+e^-}<0.93$ or 0.98,
$M_{K^+K^-e^+e^-}<3.30$ or 3.40~GeV/$c^2$, and $M^{2}_{\rm miss}\in (0.00,3.00)$ or $(-0.02,2.80)$~GeV$^2/c^4$, respectively.
For each source, the largest change of the branching fraction is less
than the statistical uncertainty after taking correlations in the samples into account, so these associated uncertainties
are neglected, as discussed in Ref.~\cite{err}.

A summary of the assigned systematic uncertainties is given in
Table~\ref{sys1}. The total relative uncertainties are determined to be 0.7\% and 1.8\% for  $D^0\to K^-e^+\nu_e$ and $D^+\to \bar K^0e^+\nu_e$, respectively, obtained by adding the individual uncertainties quadratically.

\begin{table}[htbp]
\centering
\caption{Relative systematic uncertainties (in \%) in the branching fraction measurements.
}
\small
\begin{tabular}{lcc}
  \hline
  \hline
  Source  & ${\mathcal B}_{D^{0}\to K^{-}e^{+} \nu_{e}}$ & ${\mathcal B}_{D^{+}\to \bar K^{0}e^{+} \nu_{e}}$ \\
  \hline
  $N_{D\bar D}$                &0.45  &0.4        \\
  $K^\pm$ tracking             &0.2   &...       \\
  $K^\pm$ PID                  &0.2   &...       \\
  $e^\pm$ tracking             &0.2   &0.2     \\
  $e^\pm$ PID                  &0.2   &0.2     \\
  $K_{S}^{0}$ reconstruction     &...&1.6           \\
  $E^{\rm max}_{\rm extra\,\gamma}$ and $N_{\rm extra\,\pi^{0}}$ requirements  &0.3 &0.6     \\
  ${\mathcal B}_{K^0_S\to \pi^+\pi^-}$  &...&0.1    \\
  MC model of  $D\to \bar Ke^+\nu_e$                      & 0.1 & 0.1 \\
  MC model of $D\to \bar K\pi^0 e^+\nu_e$ & 0.1 & 0.1 \\
  MC statistics                 & 0.1 & 0.2 \\ \hline
  Total & 0.7 & 1.8 \\
  \hline
  \hline
\end{tabular}
\label{sys1}
\end{table}

\begin{figure*}[htbp]
\centering
\includegraphics[width=0.48\linewidth]{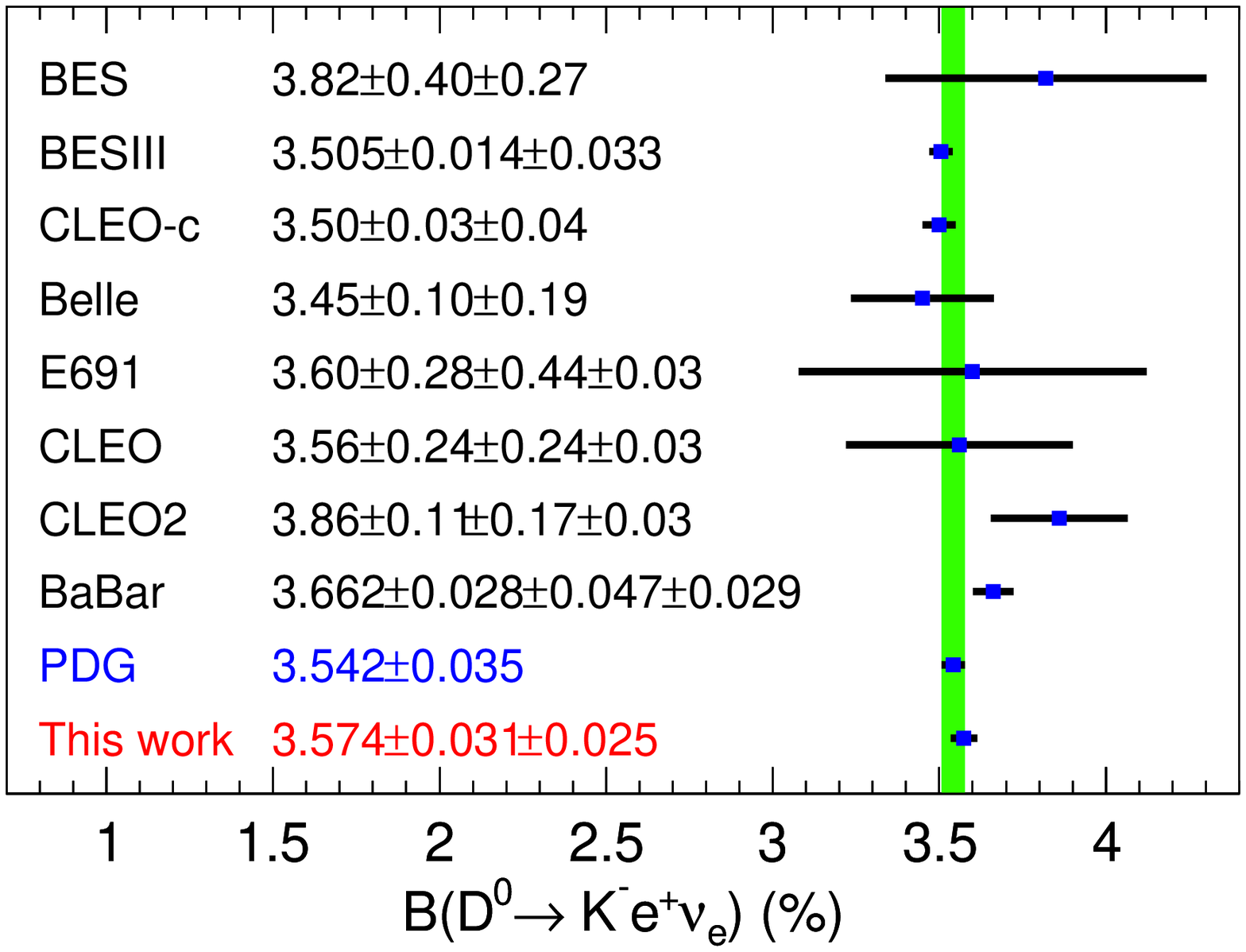}
\includegraphics[width=0.48\linewidth]{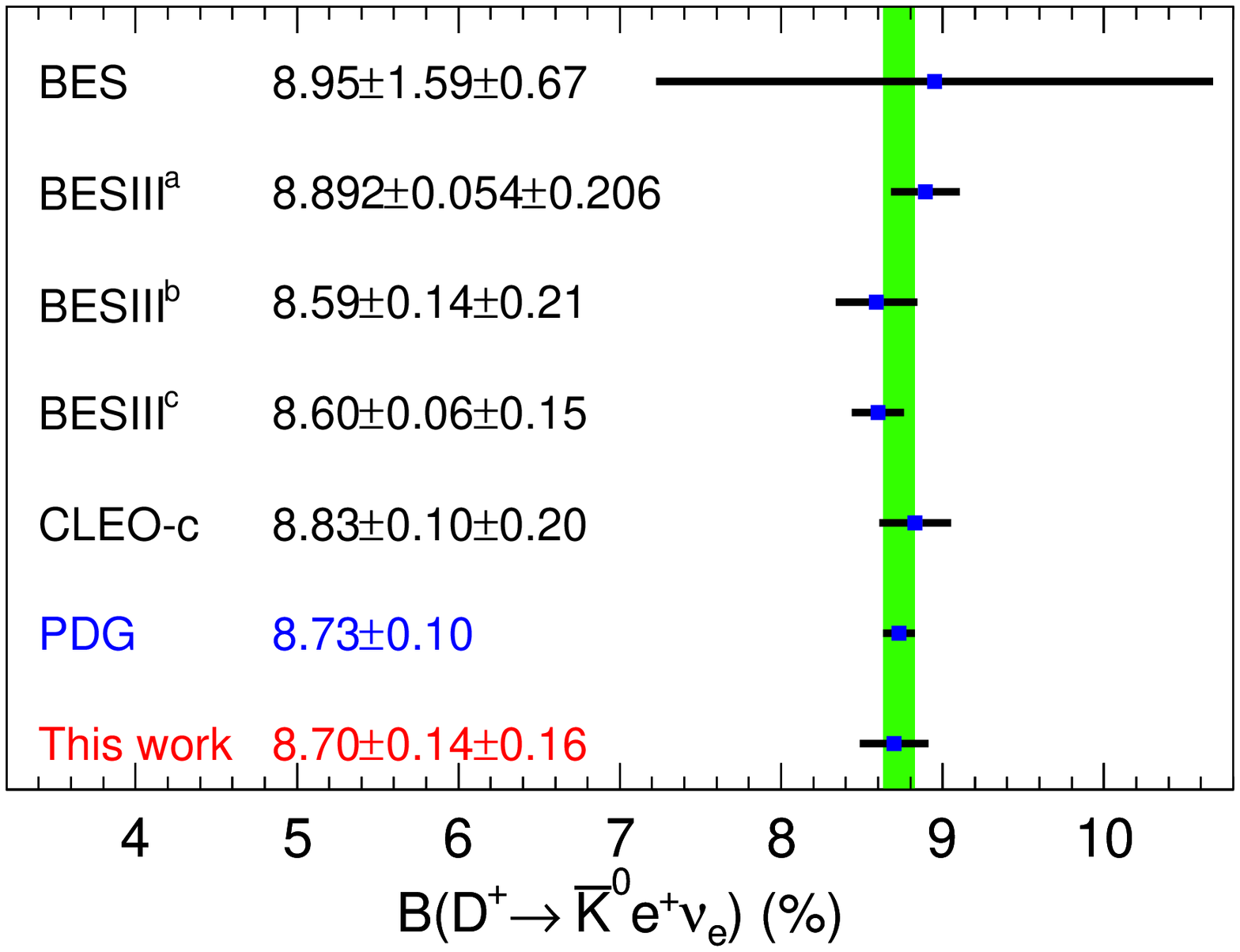}
\caption{\small
Comparison of the branching fractions from this work of (left) $D^0\to K^-e^+\nu_e$ and (right) $D^+\to \bar K^0e^+\nu_e$ to the world data. The first and second uncertainties are statistical and systematic, respectively.
For the results of E691, CLEO, CLEO2, and BaBar, the branching fractions of $D^0\to K^-e^+\nu_e$ have been
updated with the world average ${\mathcal B}_{D^0\to K^-\pi^+}$~\cite{pdg2020},
and the third uncertainty is from that of the input ${\mathcal B}_{D^0\to K^-\pi^+}$.
The branching fractions of $D^+\to \bar K^0e^+\nu_e$, labeled by BESIII$^a$, BESIII$^b$, and BESIII$^c$,
are made by using $\bar K^0\to K^0_L$, $\bar K^0\to K^0_S\to \pi^0\pi^0$,
and $\bar K^0\to K^0_S\to \pi^+\pi^-$, respectively. The green bands
correspond to the $\pm 1\sigma$ limits of the world averages.
}
\label{fig:bf}
\end{figure*}

\section{Summary}

Using an $e^+e^-$ collision data sample corresponding to an integrated luminosity of 2.93~fb$^{-1}$ collected with the BESIII detector at $\sqrt{s}=3.773$ GeV,
the absolute branching fractions of $D^{0}\to K^-e^+\nu_e$ and $D^+\to \bar K^{0}e^+\nu_e$ are determined with a new method to be
$(3.567\pm0.031_{\rm stat}\pm0.025_{\rm syst})\%$ and
$(8.68\pm0.14_{\rm stat}\pm0.16_{\rm syst})\%$.
The double-tag signal samples used in this work are independent of those adopted in our previous measurements with hadronic $\bar D^0$ or $D^-$ tags~\cite{kev,ksoev,cpc40,k8},
and
the newly obtained branching fractions have larger statistical uncertainties and smaller systematic uncertainties.
The reported branching fractions are consistent with the CCQM~\cite{CCQM}, HM$\chi$T~\cite{HMchiF}, LFQM~\cite{cheng}, and LCSR~\cite{LCSR} predictions within $2.5\sigma$.
Figure~\ref{fig:bf} gives a comparison of this measurement to the world data.
Good consistency can be seen.
Combining the obtained branching fractions for $D^0\to K^{-}e^+\nu_{e}$ and $D^{+}\to \bar K^{0}e^+\nu_{e}$ with our previous measurements
of  $D^0\to K^{-}e^+\nu_{e}$~\cite{kev} and $D^{+}\to \bar K^{0}e^+\nu_{e}$~\cite{ksoev,cpc40,k8},
we obtain
$\bar {\mathcal B}_{D^0\to K^- e^+\nu_e}=(3.531\pm0.024\pm0.017)\%$
and
$\bar {\mathcal B}_{D^+\to \bar K^0e^+\nu_e}=(8.62\pm0.07\pm0.14)\%,$
where the first error is the combined statistical and independent systematic uncertainty,
while the second error is the common systematic uncertainty.
Here, the systematic uncertainties due to tracking and PID efficiencies of $K^-$ and $e^+$,
the $K^0_S$ reconstruction and the quoted branching fraction as well as the MC model associated with form factors
are regarded as common systematic uncertainties.
Combining the re-weighted branching fractions of $D^0\to K^{-}e^+\nu_{e}$ and $D^{+}\to \bar K^{0}e^+\nu_{e}$ with the lifetimes of $D^0$ and $D^+$ mesons ($\tau_{D^0}$ and $\tau_{D^+}$), the ratio of the two decay partial widths is determined to be
\begin{equation}
\frac{\bar \Gamma_{D^0\to K^{-}e^+\nu_{e}}}{\bar \Gamma_{D^{+}\to \bar K^{0}e^+\nu_{e}}}=1.039\pm0.021.
\end{equation}
Here, the uncertainties due to the tracking and PID efficiencies of $e^\pm$ cancel, while the additional uncertainties arising from $\tau_{D^0}$ and $\tau_{D^+}$ are included.
This result supports isospin conservation in the $D^0\to K^-e^+\nu_e$ and $D^+\to \bar K^0e^+\nu_e$ decays within $1.9\sigma$.

\section{Acknowledgement}

The BESIII collaboration thanks the staff of BEPCII and the IHEP computing center for their strong support. This work is supported in part by National Key Research and Development Program of China under Contracts Nos. 2020YFA0406400, 2020YFA0406300; National Natural Science Foundation of China (NSFC) under Contracts Nos. 11775230, 11625523, 11635010, 11735014, 11822506, 11835012, 11935015, 11935016, 11935018, 11961141012; the Chinese Academy of Sciences (CAS) Large-Scale Scientific Facility Program; Joint Large-Scale Scientific Facility Funds of the NSFC and CAS under Contracts Nos. U1932102, U1732263, U1832207; CAS Key Research Program of Frontier Sciences under Contracts Nos. QYZDJ-SSW-SLH003, QYZDJ-SSW-SLH040; 100 Talents Program of CAS; INPAC and Shanghai Key Laboratory for Particle Physics and Cosmology; ERC under Contract No. 758462; European Union Horizon 2020 research and innovation programme under Marie Sklodowska-Curie grant agreement No 894790; German Research Foundation DFG under Contracts Nos. 443159800, Collaborative Research Center CRC 1044, FOR 2359, GRK 214; Istituto Nazionale di Fisica Nucleare, Italy; Ministry of Development of Turkey under Contract No. DPT2006K-120470; National Science and Technology fund; Olle Engkvist Foundation under Contract No. 200-0605; STFC (United Kingdom); The Knut and Alice Wallenberg Foundation (Sweden) under Contract No. 2016.0157; The Royal Society, UK under Contracts Nos. DH140054, DH160214; The Swedish Research Council; U. S. Department of Energy under Contracts Nos. DE-FG02-05ER41374, DE-SC-0012069.


\begin{thebibliography}{99}

\bibitem{CCQM}
M. A. Ivanov, J. G. K\"orner, J. N. Pandya, P. Santorelli, N. R. Soni, and C. T. Tran,
\href{https://link.springer.com/article/10.1007\%2Fs11467-019-0908-1}{Front. Phys. (Beijing) {\bf 14}, 64401 (2019).}

\bibitem{HMchiF}
S. Fajfer and J. F. Kamenik,
\href{https://journals.aps.org/prd/abstract/10.1103/PhysRevD.71.014020}{Phys. Rev. D {\bf 71}, 014020 (2005).}

\bibitem{cheng} H. Y. Cheng and X. W. Kang,
\href{https://doi.org/10.1140/epjc/s10052-017-5423-3}{Eur. Phys. J. C {\bf 77}, 587 (2017);}
\href{https://doi.org/10.1140/epjc/s10052-017-5170-5}{{\bf 77}, 863(E) (2017).}

\bibitem{LCSR}
Y. L. Wu, M. Zhong, and Y. B. Zuo,
\href{https://www.worldscientific.com/doi/abs/10.1142/S0217751X06033209}{Int. J. Mod. Phys. A {\bf 21}, 6125 (2006).}

\bibitem{epjc}
Y. Fang, G. Rong, H. L. Ma, and J. Y. Zhao,
\href{https://doi.org/10.1140/epjc/s10052-014-3226-3}{Eur. Phys. J. C {\bf 75}, 10 (2015).}

\bibitem{k1}
D. Besson {\it et al.} (CLEO Collaboration),
\href{https://journals.aps.org/prd/abstract/10.1103/PhysRevD.80.032005}{Phys. Rev. D {\bf 80}, 032005 (2009).}

\bibitem{kev}
M. Ablikim {\it et al.} (BESIII Collaboration),
\href{https://journals.aps.org/prd/abstract/10.1103/PhysRevD.92.072012}{Phys. Rev. D {\bf 92}, 072012 (2015).}

\bibitem{ksoev}
M. Ablikim {\it et al.} (BESIII Collaboration),
\href{https://journals.aps.org/prd/abstract/10.1103/PhysRevD.96.012002}{Phys. Rev. D {\bf 96}, 012002 (2017).}

\bibitem{pdg2020}
P. A. Zyla {\it et al.} (Particle Data Group),
\href{http://pdglivetest.lbl.gov/Viewer.action}{Prog. Theor. Exp. Phys. {\bf 2020}, 083C01 (2020).}

\bibitem{bes2-kev}
M. Ablikim {\it et al.} (BES Collaboration),
\href{https://doi.org/10.1016/j.physletb.2004.07.004}{Phys. Lett. B {\bf 597}, 39 (2004).}

\bibitem{bes2-ksev}
M. Ablikim {\it et al.} (BES Collaboration),
\href{https://doi.org/10.1016/j.physletb.2004.12.040}{Phys. Lett. B {\bf 608}, 24 (2005).}

\bibitem{k2}
L. Widhalm {\it et al.} (Belle Collaboration),
\href{https://journals.aps.org/prl/abstract/10.1103/PhysRevLett.97.061804}{Phys. Rev. Lett. {\bf 97}, 061804 (2006).}

\bibitem{cpc40}
M. Ablikim {\it et al.} (BESIII Collaboration),
\href{https://iopscience.iop.org/article/10.1088/1674-1137/40/11/113001}{Chin. Phys. C {\bf 40}, 113001 (2016).}

\bibitem{k8}
M. Ablikim {\it et al.} (BESIII Collaboration),
\href{https://journals.aps.org/prd/abstract/10.1103/PhysRevD.92.112008}{Phys. Rev. D {\bf 92}, 112008 (2015).}

\bibitem{k6}
J. C. Anjos {\it et al.} (E691 Collaboration),
\href{https://journals.aps.org/prl/abstract/10.1103/PhysRevLett.62.1587}{Phys. Rev. Lett. {\bf 62}, 1587 (1989).}

\bibitem{k5}
G. Crawford {\it et al.} (CLEO Collaboration),
\href{https://journals.aps.org/prd/abstract/10.1103/PhysRevD.44.3394}{Phys. Rev. D {\bf 44}, 3394 (1991).}

\bibitem{k4}
A. Bean {\it et al.} (CLEO Collaboration),
\href{https://doi.org/10.1016/0370-2693(93)91385-Z}{Phys. Lett. B {\bf 317}, 647 (1993).}

\bibitem{k3}
B. Aubert {\it et al.} (BaBar Collaboration),
\href{https://journals.aps.org/prd/abstract/10.1103/PhysRevD.76.052005}{Phys. Rev. D {\bf 76}, 052005 (2007).}


\bibitem{NDD} M. Ablikim {\it et al}. (BESIII Collaboration),
\href{https://iopscience.iop.org/article/10.1088/1674-1137/42/8/083001}{Chin. Phys. C {\bf 42}, 083001 (2018).}

\bibitem{BESIII}
M. Ablikim {\it et al.} (BESIII Collaboration),
\href{https://doi.org/10.1016/j.nima.2009.12.050}{Nucl. Instrum. Meth. A {\bf 614}, 345 (2010).}

\bibitem{Yu:IPAC2016-TUYA01}
C.~H.~Yu {\it et al.},
\href{https://doi.org/10.18429/JACoW-IPAC2016-TUYA01}{Proceedings of IPAC2016, Busan, Korea, 2016.}

\bibitem{cpc41}
M. Ablikim {\it et al.} (BESIII Collaboration),
\href{http://hepnp.ihep.ac.cn/en/article/doi/10.1088/1674-1137/44/4/040001}{Chin. Phys. C {\bf 44}, 040001 (2020).}

\bibitem{geant4}
S. Agostinelli {\it et al.} (GEANT4 Collaboration),
\href{https://doi.org/10.1016/S0168-9002(03)01368-8}{Nucl. Instrum. Meth. A {\bf 506}, 250 (2003).}

\bibitem{kkmc}
S. Jadach, B. F. L. Ward, and Z. Was,
\href{https://linkinghub.elsevier.com/retrieve/pii/S0010465500000485}{ Comp. Phys. Commu. {\bf 130}, 260 (2000);} \href{https://journals.aps.org/prd/abstract/10.1103/PhysRevD.63.113009}{Phys. Rev. D {\bf 63}, 113009 (2001).}

\bibitem{evtgen}
D.~J.~Lange,
\href{https://doi.org/10.1016/S0168-9002(01)00089-4} {Nucl. Instrum. Meth. A {\bf 462}, 152 (2001);}
R.~G.~Ping,
\href{https://doi.org/10.1088/1674-1137/32/8/001}{Chin. Phys. C {\bf 32}, 599 (2008).}

\bibitem{lundcharm}
J. C. Chen, G. S. Huang, X. R. Qi, D. H. Zhang, and Y. S. Zhu,
\href{https://journals.aps.org/prd/abstract/10.1103/PhysRevD.62.034003}{Phys. Rev. D {\bf 62}, 034003 (2000).}

\bibitem{photos}
E.~Richter-Was,
\href{https://doi.org/10.1016/0370-2693(93)90062-M`}{Phys. Lett. B {\bf 303}, 163 (1993).}

\bibitem{MPM} D. Becirevic and A. B. Kaidalov,
\href{https://doi.org/10.1016/S0370-2693(00)00290-2}{Phys. Lett. B {\bf 478}, 417 (2000).}

\bibitem{epjc76}
M. Ablikim {\it et al.} (BESIII Collaboration),
\href{https://doi.org/10.1140/epjc/s10052-016-4198-2}{Eur. Phys. J. C {\bf 76}, 369 (2016).}

\bibitem{bes3-pimuv}
M. Ablikim {\it et al.} (BESIII Collaboration),
\href{https://journals.aps.org/prl/abstract/10.1103/PhysRevLett.121.171803}{Phys. Rev. Lett. {\bf 121}, 171803 (2018).}

\bibitem{bes3-Dp-K1ev}
M. Ablikim {\it et al.} (BESIII Collaboration),
\href{https://journals.aps.org/prl/abstract/10.1103/PhysRevLett.123.231801}{Phys. Rev. Lett. {\bf 123}, 231801 (2019).}

\bibitem{bes3-etaetapi}
M. Ablikim {\it et al.} (BESIII Collaboration),
\href{https://journals.aps.org/prd/pdf/10.1103/PhysRevD.101.052009}{Phys. Rev. D {\bf 101}, 052009 (2020).}

\bibitem{bes3-omegamuv}
M. Ablikim {\it et al.} (BESIII Collaboration),
\href{https://journals.aps.org/prd/pdf/10.1103/PhysRevD.101.072005}{Phys. Rev. D {\bf 101}, 072005 (2020).}

\bibitem{bes3-etamuv}
M. Ablikim {\it et al.} (BESIII Collaboration),
\href{https://journals.aps.org/prl/abstract/10.1103/PhysRevLett.124.231801}{Phys. Rev. Lett. {\bf 124}, 231801 (2020).}

\bibitem{bes3-etaX}
M. Ablikim {\it et al.} (BESIII Collaboration),
\href{https://journals.aps.org/prl/abstract/10.1103/PhysRevLett.124.241803}{Phys. Rev. Lett. {\bf 124}, 241803 (2020).}

\bibitem{bes3-DCS-Dp-K3pi}
M. Ablikim {\it et al.} (BESIII Collaboration),
\href{https://journals.aps.org/prl/abstract/10.1103/PhysRevLett.125.141802}{Phys. Rev. Lett. {\bf 125}, 141802 (2020).}

\bibitem{bes3-D-KKpipi}
M. Ablikim {\it et al.} (BESIII Collaboration),
\href{https://journals.aps.org/prd/abstract/10.1103/PhysRevD.102.052006}{Phys. Rev. D {\bf 102}, 052006 (2020).}

\bibitem{bes3-D-b1enu}
M. Ablikim {\it et al}. (BESIII Collaboration),
\href{https://journals.aps.org/prd/abstract/10.1103/PhysRevD.102.112005}{Phys. Rev. D {\bf 102}, 112005 (2020).}

\bibitem{D0_kpiev}
M. Ablikim {\it et al.} (BESIII Collaboration),
\href{https://journals.aps.org/prd/pdf/10.1103/PhysRevD.94.032001}{Phys. Rev. D {\bf 94}, 032001 (2016).}

\bibitem{Dp_kpiev}
M. Ablikim {\it et al.} (BESIII Collaboration),
\href{https://journals.aps.org/prd/abstract/10.1103/PhysRevD.99.011103}{Phys. Rev. D {\bf 99}, 011103 (2019).}

\bibitem{err}
R. Barlow, \href{https://arxiv.org/abs/hep-ex/0207026}{arXiv:hep-ex/0207026.}

\end{thebibliography}
\end{document}